\documentclass[]{spie}  %>>> use for US letter paper
\usepackage{amsmath}
\usepackage[]{graphicx}
\usepackage{cite}
\usepackage{float}
\newfloat{Video}{t}{}

\title{Woofer-tweeter deformable mirror control for closed-loop adaptive optics: theory and practice} 

\author{Donald Gavel\supit{a} and Andrew Norton\supit{b}
\skiplinehalf
\supit{a}Universiy fo California Observatories, 1156 High Street, Santa Cruz, CA, USA; \\
\supit{b}Lockheed Martin Space Systems, Palo Alto, CA, USA
}

\authorinfo{Further author information: (Send correspondence to DG)\\DG: E-mail: gavel@ucolick.org, Telephone: 1-831-459-5464\\ AN: E-mail: apnorton@gmail.com}
\begin{document} 
  \maketitle 

%%%%%%%%%%%%%%%%%%%%%%%%%%%%%%%%%%%%%%%%%%%%%%%%%%%%%%%%%%%%% 
\begin{abstract}
Deformable mirrors with very high order correction generally have smaller dynamic range of motion than what is required to correct seeing over large aperture telescopes.  As a result, systems will need to have an architecture that employs two deformable mirrors in series, one for the low-order but large excursion parts of the wavefront and one for the finer and smaller excursion components. The closed-loop control challenge is to a) keep the overall system stable, b) avoid the two mirrors using control energy to cancel each other’s correction, c) resolve actuator saturations stably, d) assure that on average the mirrors are each correcting their assigned region of spatial frequency space. We present the control architecture and techniques for assuring that it is linear and stable according to the above criteria. We derived the analytic forms for stability and performance and show results from simulations and on-sky testing using the new ShaneAO system on the Lick 3-meter telescope. \end{abstract}

%>>>> Include a list of keywords after the abstract 

\keywords{adaptive optics, woofer-tweeter, stability, wavefront control, deformable mirrors}

%%%%%%%%%%%%%%%%%%%%%%%%%%%%%%%%%%%%%%%%%%%%%%%%%%%%%%%%%%%%%
\section{INTRODUCTION}
\label{sec:intro}  % \label{} allows reference to this section

Every deformable mirror, depending on the number and layou of actuators, and each individual actuator's influence function, can address a limited subspace of the overall space of possible wavefront aberrations. Adaptive optics controllers attempt to find linear combinations of deformable mirror actuators that minimize the wavefront error after reflection, with the result that a planar wavefront enters the science instrument.

New adaptive optics systems are demanding a better fit and demanding it over larger apertures. The consequence is DM need more actuators and a larger stroke for each actuator. The later is because the typical atmospheric aberration has peak to valley phase excursions that increase with telescope diameter, when referenced to the mean wavefront over the aperture. Unfortunately, deformable mirror technology can only provide large stroke with limited numbers of actuators or large numbers of actuators with limited stroke. Therefore we explore the notion of a ``woofer-tweeter" system of two mirrors, one to handle lower order aberrations at large stroke and one to correct the finer scale and lower deflection aberrations. Fortunately, Kolmogorov atmospheric statistics have a power law ($\kappa^{-11/3}$) that favors the larger excursions of the wavefront to be at lower spatial frequencies. In a dual-DM system, it is important that the control law sorts out the orders so that the lower orders tend to go to the woofer and the higher orders to the tweeter. Furthermore, large stroke motions (of the woofer) can have a limited slew rate depending on the mirror design. In this case, we want the low-order excursions to temporarily go to the tweeter, but have the woofer catch up before the tweeter saturates.

Woofer-tweeter control architectures have been studied for application on AO systems for some time.\cite{Hampton2006, Lavigne2008, Farrell2008, Correia2012, Zou2012}  Recently, an AO system with this control architecture in Fourier space has been successfully deployed on sky with the Gemini Planet Imager\cite{Lavigne2007}. In the present paper we describe a technique that is mathematically general with respect to mode spaces and is applicable to when the woofer and tweeter have different temporal responses. We prove overall stability and show that it meets the objective of sorting modes correctly between woofer and tweeter. This architecture has now been demonstrated on sky with the new ShaneAO system.\cite{Gavel2011, Kupke2012}

Actuator saturation is the enemy of adaptive optics. Not only does it allow aberration to work its way into the science beam and degrade it, it can  cause instability in the control loop itself, further degrading the science image quality. Even if one uses the best known stable ``anti-reset-windup" techniques \cite{Astrom1989, Wittenmark1989}, complex AO controllers still tend to get stuck in saturation for periods much longer than the original  bad seeing burst that initiated it. The best approach is to keep the system in a linear regime by avoiding saturation altogether . This is a major advantage of the woofer-tweeter architecture.

\section{MATHEMATICAL PRELIMINARIES}
\label{sec:mathprelim}

To model how the woofer and tweeter interact, we use the techniques of linear algebra applied to both the finite dimensional spaces of the control vectors and also to the infinite dimensional (Hilbert) spaces spanned by wavefront phases and DM response functions.

\subsection{Modal spaces} 
\label{sec:modes}

We assign a set of modes $b_i(x)$ to each DM, such that the modes can fit an arbitrary phase $\phi(x)$ with some error: 
\begin{equation}
\phi_b(x) = \sum_i {c_i b_i(x)} \quad\quad e_\phi = \phi (x)-\phi_b(x).
\end{equation}
Here, $\left\{c_i, b_i(x) \right\}$ is any vector space. We make no assumption about these modal basis functions other they are functions - single-valued across the face of the mirror. The modes do not have to be orthogonal, normalized, or even linearly independent. They can be discontinuous (e.g. a segmented DM). The mode set can mix pieces of ``familiar" mode sets like Zernike, Fourier, etc. Solutions $\phi_b(x)$ are restricted to the subspace spanned by the basis functions.

The Shack-Hartmann wavefront sensor responds to the wavefront as
\begin{equation}
s_j = \int {w_j(x) \nabla \phi(x) dx} + n_j; \quad i \in \{subaps\} 
\end{equation}
where $w_j(s)$ represents weighting functions defining the subapertures. The sensor's readings are related to the mode coefficients by
\begin{equation}
{\bf s} = {\bf H c}+e_{s}
\end{equation}
where the entries of the $\bf H$ matrix are found by integrating $\nabla b_i(x)$ with $w_j(x)$. In practice, the ${\bf H}$ matrix is assembled experimentally by sending modal command vectors to the DM and measuring the response on the wavefront sensor. The error $e_s$ is the combination of the mearuement noise $n_j$ and the integrals of $e_\phi$ with $w_j(x)$.

Finally, we assume that the actuators on the deformable mirrors can reproduce the basis functions, again with some error:
\begin{equation}
\phi_b(x) = \sum_i {a_i r_i(x)} + e_{fit}
\end{equation}
where $r_i(x)$ are the actuator influence functions. The fit error $e_{fit}$ can be very small if we, say, use natural modes of the deformabable mirror, but we leave it there to remind us that the mirror may still have some residual nonlinearity or other inability to fit modes exactly. There is no assumption that the Hilbert space spanned by the actuator response functions is spanned by the modal basis functions. Only the converse is assumed, that the modal space is spanned by the actuator responses. In fact it is sometimes convenient that the modal space is restricted to fewer degrees of freedom than the mirror can produce, for example, if we wish to use only a few or the low-order modes of the woofer DM in the woofer-tweeter control scheme.

The actuator command vector is related to the mode coefficients by
\begin{equation}\label{eq:A}
{\bf a} = {\bf A c}
\end{equation}
where the $i$'th column of ${\bf A}$ is the vector of actuator commands that produce mode $b_i(x)$ on the mirror. 

It should be stressed that these equation are in a very general form for linear models of the DM(s) and wavefront sensor behavior. The basis functions are arbitrary, and somewhat artificially introduced here, but they are intended to give us flexibility down the road so that the woofer-tweeter wavefront control solution is applicable to any of the favored basis sets used in the AO community or can be tuned to the particular kinds of DMs or other situations. Of course there is practical advantage to choosing modes that can be well fit with some combination of actuator commands; in the end this will reduce overall fitting error. In our application case, the ShaneAO adaptive optics system, we exploited the generality to allow our modes to be the natural mirror modes for each of woofer and tweeter, and then used the richness of the tweeter mode space to fit woofer modes with it using the process described in section~\ref{subsec:fitWooferToTweeter}.

\subsection{Reconstructor} 
\label{sec:reconstructor}

The reconstructor strives to find the mode coefficients given the sensor readings:
\begin{gather} \label{eq:R}
{\bf c} = {\bf H}^\dagger {\bf s} \\ \label{eq:H}
{\bf H}^\dagger = ({\bf Q} + {\bf H}^T {\bf P H})^{-1} {\bf H}^T {\bf P}  
\end{gather}
where $\bf P$ and $\bf Q$ are weighting and reqularization matrices respectively. This general form for the linear reconstructor depicts, through appropriate choise of $\bf P$ and $\bf Q$, all the popular variants of vector-matrix reconstructor including weighted least-squares, minimum variance, and maximum a posteriori (MAP) estimation. It can also incorporate the actuator penalty methods for suppressing waffle and other unwanted DM modes \cite{Gavel2003}.

WIth the mode coefficients provided by the reconstructor, the actuators are set using \eqref{eq:A}, thus
\begin{equation}\label{eq:R2}
{\bf a} = {\bf R} {\bf s}; \qquad {\bf R} = {\bf A} {\bf H}^\dagger.
\end{equation}

\begin{figure}
   \begin{center}
   \begin{tabular}{c}
   \includegraphics[height=3.5in]{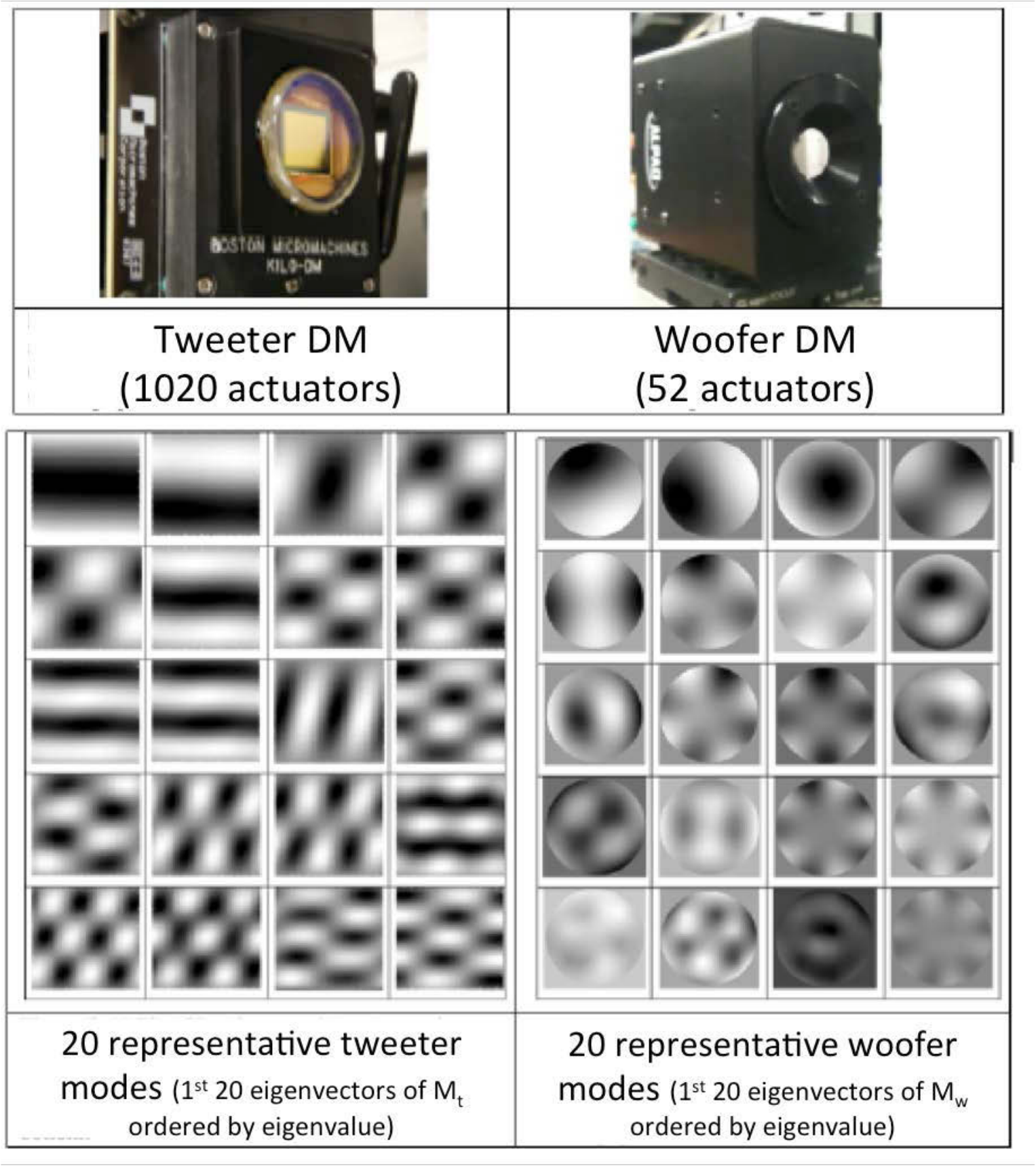}
   \end{tabular}
   \end{center}
   \caption[example] 
%>>>> use \label inside caption to get Fig. number with \ref{}
   { \label{fig:modesets} 
Woofer and tweeter deformable mirrors from the ShaneAO system along with samples from the natural mode sets of each.\cite{Norton2013}}
   \end{figure}

Examples of woofer and tweeter mode spaces from the ShaneAO adaptive optics system are shown in Figure~\ref{fig:modesets}. This system uses a 1024-actuator tweeter DM and a 52-actuator woofer DM. The mode sets used are composed of natural mirror modes, i.e. those defined by the eigenvectors of
\begin{equation}
M_{ij} = \int {r_i(x) r_j(x) dx}.
\end{equation}

%-------------
%-------------

%%%%%%%%%%%%%%%%%%%%%%%%%%%%%%%%%%%%%%%%%%%%%%%%%%%%%%%%%%%%%
\section{WOOFER AND TWEETER MODE SPACES} 
%\subsection{Woofer and tweeter mode spaces}

The woofer and tweeter respond to linear combinations of their mode sets according to 
\begin{equation}\label{eq:Phase}
\phi_w(x) = \sum_i {c_{w_i} b_{w_i}(x)} \quad\quad \phi_t(x) = \sum_i {c_{t_i} b_{t_i}(x)}.
\end{equation}
The coefficients that provide the least-squares fits to a given wavefront $\phi(x)$ are
\begin{equation}
{\bf c}_w = {\bf M}_w \left[ \int b_{w_i}(x) \phi(x) dx \right]; \quad \quad {\bf c}_t = {\bf M}_t \left[ \int b_{ti}(x) \phi(x) dx \right]
\end{equation}
where
\begin{equation}\label{eq:M}
{\bf M}_w = \left [ \int b_{w_i}(x) b_{w_j}(x) dx \right ]^{-1}; \quad \quad {\bf M}_t = \left [ \int b_{t_i}(x) b_{t_j}(x) dx \right ]^{-1}.
\end{equation}
Accordingly, a wavefront phase defined by actuator motions on the tweeter can be projected to the woofer by:
\begin{equation}\label{eq:CrossFit}
{\bf c}_w = {\bf M}_w \left[ \int b_{w_i}(x) \sum_j c_{t_j} b_{t_j}(x) dx \right ] = {\bf M}_w {\bf C}_{wt} {\bf c}_t
\end{equation}
where
\begin{equation}\label{eq:C}
{\bf C}_{wt} = \left [ \int b_{w_i}(x) b_{t_j}(x) dx \right ].
\end{equation}

\subsection{Hilbert Matrices}

To simplify the notation so that from here on we can avoid the distracting integral signs, we re-write the various functions of $x$ as infinite-dimensional vectors, and indexed functions of $x$ as infinite-dimensional, along one side, matrices. Such infinite dimensional vector spaces are known as Hilbert spaces \cite{Hilbert1927}. Equation \eqref{eq:Phase} becomes:
\begin{equation}
{\boldsymbol \phi}_t = {\bf B}_t^T {\bf c}_t \quad {\boldsymbol \phi}_w = {\bf B}_w^T {\bf c}_w
\end{equation}
and equations \eqref{eq:M} and \eqref{eq:C} become
\begin{equation}\label{eq:HilbertNotation}
{\bf M}_w = \left [{\bf B}_w {\bf B}_w^T \right] ^{-1} \quad {\bf M}_t = \left [ {\bf B}_t {\bf B}_t^T \right] ^{-1} \quad {\bf C}_{wt} = {\bf B}_w {\bf B}_t^T
\end{equation}
where
\begin{equation}
{\bf B}_t = \left [ \begin {matrix} b_{t_0}(x) \\ b_{t_1}(x) \\ ...\end{matrix} \right ].
\end{equation}

\subsection {Least Squares Fits of Woofer to Tweeter Modes} \label{subsec:fitWooferToTweeter}
If the tweeter can faithfully produce some of the woofer modes then this will form an overlapping subspace that both the woofer and the tweeter can address. This will form the basis for a control loop that offloads low-order corrections from the tweeter to the woofer.

When the Hilbert space notation \eqref{eq:HilbertNotation} is substituted into the formulas for cross-fits between the mode spaces \eqref{eq:CrossFit}, pseudo-inverses of ${\bf B}_w$ and ${\bf B}_t$ appear.
\begin{equation}
{\bf M}_w {\bf C}_{wt} = \left [ {\bf B}_w {\bf B}_w^T \right ]^{-1} {\bf B}_w {\bf B}_t^T = {\bf B}_w^\dagger {\bf B}_t^T ;  \qquad
{\bf M}_t {\bf C}_{wt}^T = \left [ {\bf B}_t {\bf B}_t^T \right]^{-1} {\bf B}_t {\bf B}_w^T = {\bf B}_t^\dagger {\bf B}_w^T
\end{equation}
Let
\begin{equation}
{\boldsymbol \phi} = {\bf B}_t {\bf c}_t = {\bf B}_w {\bf c}_w
\end{equation}
be in the overlapping subspace, the subspace of possible $\phi(x)$ where $ \phi(x)$ is reasonably well fit by either the woofer or the tweeter. Then
\begin{equation}
{\bf M}_t {\bf C}_{wt}^T {\bf M}_w {\bf C}_{wt} {\bf c}_t 
= {\bf B}_t^\dagger {\bf B}_w^T {\bf B}_w^\dagger {\bf B}_t^T {\bf c}_t
= {\bf B}_t^\dagger {\bf B}_w^T {\bf B}_w^\dagger {\bf B}_w^T {\bf c}_w,
\end{equation}
and since ${\bf B}_w^\dagger {\bf B}_w^T = I$ and ${\bf B}_t^\dagger {\bf B}_t^T = {\bf I}$ this collapses to
\begin{equation}
{\bf M}_t {\bf C}_{wt}^T {\bf M}_w {\bf C}_{wt} {\bf c}_t ={\bf B}_t^\dagger {\bf B}_w^T {\bf c}_w = {\bf B}_t^\dagger {\bf B}_t^T {\bf c}_t = {\bf c}_t,
\end{equation}
which proves that projecting any $\boldsymbol \phi$ that is in the overlapping mode space from the tweeter to the woofer, followed by projecting  back on to the tweeter, is an identity operation.
 
%%%%%%%%%%%%%%%%%%%%%%%%%%%%%%%%%%%%%%%%%%%%%%%%%%%%%%%%%%%%%
\section{WOOER-TWEETER CONTROL ARCHITECTURE} 

With this idea of an overlapping mode space in mind, we construct a control framework where the wavefront correction is split between woofer and tweeter. Nominally, the split is such that overlapping modes go only to the woofer and are subtracted from the tweeter commands. The cross-fits described in the prior mathematical sections are used to ensure that the two deformable mirrors are controlling orthogonal subspace, i.e. do not ``fight'' with each other.

\begin{figure}
   \begin{center}
   \begin{tabular}{c}
   \includegraphics[height=2.3in]{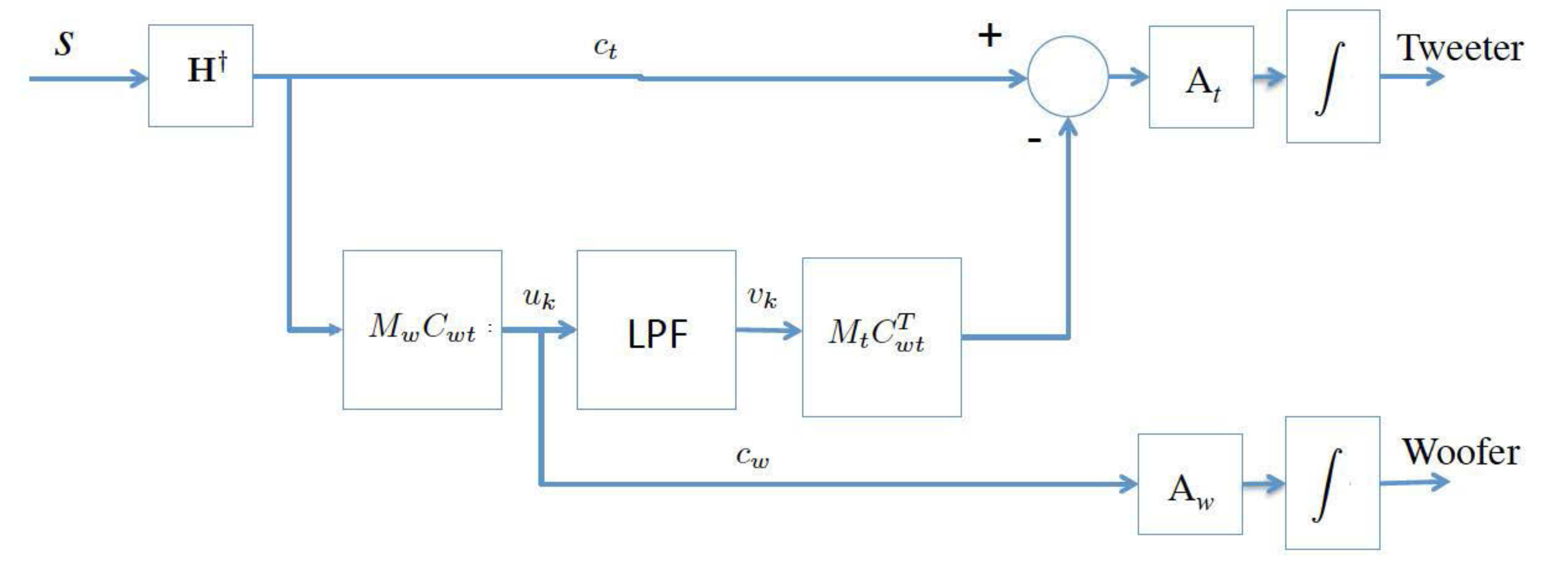}
   \end{tabular}
   \end{center}
   \caption[example] 
%>>>> use \label inside caption to get Fig. number with \ref{}
   { \label{fig:blockDiagram} 
Woofer-tweeter control architecture. LPF is a low-pass filter, designed with a time constant matching the woofer response. A rapid change in low-order mode initially appears on the tweeter. As the woofer slowly responds to it, the LPF output then acts to suppress this mode from the tweeter.}
   \end{figure}

However, woofer DMs tend to respond slower than tweeter DMs. We propose to accont for the speed difference by allowing the overlapping woofer modes to temporaily go on the tweeter but then let them bleed off of the tweeter on a time scale comparable to the woofer's temporal response. The architecture is shown in Figure~\ref{fig:blockDiagram}. Incomming reconstructed wavefront is sent entirely to the tweeter and also projected to woofer mode coefficients. The resulting woofer space solution is sent both directly to the woofer and the mode coefficinets to a low-pass filter (LPF). The low-pass filtered ouput, now projected back on to the tweeter space, is then subtracted from the tweeter commands.

%-------------
%-------------

The controller takes wavefront sensor signal ${\bf s}$ and generates mirror actuator command vectors ${\bf a}$. The diagram in Figure~\ref{fig:blockDiagram} depicts closed-loop operation, where integrators
(  $ \int = 1 / \left(1 - z^{-1}\right) $)
accumulate delta-commands in response to wavefront sensor measurements which, in closed loop, are atmospheric phase minus the previous DM corrections.

The simplest low-pass filter is single-pole:
\begin{equation} \label{eq:filterState}
v_k = \alpha v_{k-1} + (1-\alpha)u_{k-1}; \quad \left | \alpha \right | < 1
\end{equation}
which has a $z$-transform transfer function
\begin{equation}
H_L(z^{-1}) = {{ z^{-1} (1-\alpha) } \over {1 - \alpha z^{-1}}}.
\end{equation}
The filter has a pole at $z=\alpha$ and therefore a $1/e$ time constant of $\tau = T / \ln{\alpha^{-1}}$, where $T$ is the controller's sample period.

Substituting for $u$ in \eqref{eq:filterState} the negative-feedback control from sensor to filter is, in the state-space form:
\begin{equation} \label{eq:filterState2}
{\bf v}_k = \alpha {\bf v}_{k-1} - (1-\alpha){\bf M}_w {\bf C}_{wt} {\bf H}^\dagger {\bf s}_k
\end{equation}
and the controls to the woofer and tweeter are:
\begin{align} \label{eq:wooferState}
{\bf a}_{w_k} &= {\bf a}_{w_{k-1}} - {\bf A}_w {\bf M}_w {\bf C}_{wt} {\bf H}^\dagger {\bf s}_k \\ \label{eq:tweeterState}
{\bf a}_{t_k} &= {\bf a}_{t_{k-1}} - {\bf R} {\bf s}_k - {\bf A}_t {\bf M}_t {\bf C}_{wt}^T {\bf v}_{k-1} .
\end{align}

For the purposes of our subsequent stability analysis, it is convenient to describe the control system in a block-matrix form. Combining \eqref{eq:filterState2} -- \eqref{eq:tweeterState} we get
\begin{equation} \label{eq:controller}
\left[ \begin{matrix}{\bf a}_t \\ {\bf a}_w \\ {\bf v} \end{matrix} \right]_k = 
\left[ \begin{matrix}{\bf I} & {\bf 0} &  - {\bf A}_t {\bf M}_t {\bf C}_{wt}^T \\
                       {\bf 0} & {\bf I} & {\bf 0} \\
                       {\bf 0} & {\bf 0} & \alpha \end{matrix} \right]
\left[ \begin{matrix}{\bf a}_t \\ {\bf a}_w \\ {\bf v} \end{matrix} \right]_{k-1} - 
\left[ \begin{matrix}{\bf R}  \\ 
                  {\bf A}_w {\bf M}_w {\bf C}_{wt} {\bf H}^\dagger \\
                 (1-\alpha){\bf M}_w {\bf C}_{wt} {\bf H}^\dagger\end{matrix} \right ] {\bf s}_k.
\end{equation}

\section {STABILITY AND PERFORMANCE}
There are two major questions to be answered: 1) is the closed loop system stable, with some margin of stability that increases as a parameter such as closed-loop gain is reduced, and 2) is the system performing as desired. In our case, performance is defined not only by good fit to wavefront aberration, but also by whether it is accomplishing the objective of putting most of the excursion range on the woofer, while well-fitting small high-frequency deviations with the tweeter.

To analyze closed-loop stability, we must add the mathematical models for how the wavefront measurements ${\bf s}_k$ are affected by  actuator commands, ${\bf a}_{w_k}$ and ${\bf a}_{t_k}$. The woofer and tweeter each contribute additively to the wavefront correction:
\begin{equation} \label{eq:wfs}
{\bf s}_k = {\bf H}_t {\bf A}_t^\dagger {\bf a}_{t_{k-1}} + {\bf H}_w {\bf A}_w^\dagger \bar {\bf a}_{w_{k-1}}
\end{equation}
where ${\bf H}_w$ describes how the woofer modes affect the measurements and ${\bf H}_t = {\bf H}$ describes how the tweeter modes affect the measurements. ${\bf A}^\dagger = ({\bf A}^T {\bf A})^{-1} {\bf A}^T$ describe how the actuator vectors map to mode coefficients. Note that actuator command vectors are restricted to be in the subspace spanned by the modes by equation~\eqref{eq:R2}, so ${\bf A}^\dagger$ and ${\bf A}$ can be used to map uniquely between mode coefficient and actuator command vectors.

Since the woofer does not immediately respond to commands ${\bf a}_{w_k}$, we use $\bar {\bf a}_{w_k}$ to denote the actual value of the woofer actuators at any given time step $k$. The woofer has a finite time response, which is modeled as:
\begin{equation} \label{eq:wooferLag}
\bar {\bf a}_{w_k} = \beta \bar {\bf a}_{w_{k-1}} + (1-\beta) {\bf a}_{w_{k-1}}
\end{equation}
where ${\bf a}_w$ is the woofer command and $\beta$ is related to the time constant of the woofer response ($\tau = {T / {\ln \beta^{-1}}}$). It is assumed that the tweeter responds instantaneously.

Combining \eqref{eq:controller}, \eqref{eq:wfs}, and \eqref{eq:wooferLag} we have the closed-loop matrix equation
\begin{equation}
\left[ \begin{matrix}{\bf a}_t \\ {\bf a}_w \\ {\bf v} \\ \bar {\bf a}_w\end{matrix} \right]_k = 
\left[ \begin{matrix}{\bf I} &   {\bf  0}      &  - {\bf A}_t {\bf M}_t {\bf C}_{wt}^T  & {\bf 0}       \\
                     {\bf 0} &   {\bf I}      &            {\bf 0}                   & {\bf 0}        \\
                     {\bf 0} &    {\bf 0}     &       \alpha {\bf I}              & {\bf 0}       \\ 
                     {\bf 0} &    (1-\beta){\bf I}     &           {\bf 0}        & \beta {\bf I} \end{matrix}    \right] 
\left[ \begin{matrix}{\bf a}_t \\ {\bf a}_w \\ {\bf v} \\ \bar {\bf a}_w\end{matrix} \right]_{k-1} - 
\left[ \begin{matrix}{\bf A}_t {\bf H}_t^\dagger \\
                      {\bf A}_w {\bf M}_w {\bf C}_{wt} {\bf H}_t^\dagger \\ 
                     (1-\alpha){\bf M}_w {\bf C}_{wt} {\bf H}_t^\dagger 
                      \\ {\bf 0}\end{matrix} \right ]
\left[ \begin{matrix}{\bf H}_t {\bf A}_t^\dagger & {\bf 0} & {\bf 0} & {\bf H}_w {\bf A}_w^\dagger \end{matrix} \right]
\left[ \begin{matrix}{\bf a}_t \\ {\bf a}_w \\ {\bf v} \\ \bar {\bf a}_w\end{matrix} \right]_{k-1}
\end{equation}
or
\begin{equation}
\left[ \begin{matrix}{\bf a}_t \\ {\bf a}_w \\ {\bf v} \\ \bar {\bf a}_w\end{matrix} \right]_k =
\left [\begin{matrix}
{\bf I} - {\bf A}_t {\bf H}_t^\dagger {\bf H}_t {\bf A}_t^\dagger &        {\bf  }0          &            -  {\bf A}_t {\bf M}_t {\bf C}_{wt}^T       &     -{\bf A}_t {\bf H}_t^\dagger {\bf H}_w {\bf A}_w^\dagger  \\
- {\bf A}_w {\bf M}_w {\bf C}_{wt} {\bf H}_t^\dagger {\bf H}_t {\bf A}_t^\dagger    & {\bf I}  &   {\bf 0}       &   - {\bf A}_w {\bf M}_w {\bf C}_{wt} {\bf H}_t^\dagger {\bf H}_w {\bf A}_w^\dagger   \\
-(1-\alpha) {\bf M}_w {\bf C}_{wt} {\bf H}_t^\dagger {\bf H}_t {\bf A}_t^\dagger & {\bf 0}        & \alpha {\bf I}    & -(1-\alpha) {\bf M}_w {\bf C}_{wt} {\bf H}_t^\dagger {\bf H}_w {\bf A}_w^\dagger  \\
{\bf 0} & (1-\beta) {\bf I} & {\bf 0} & \beta {\bf I} \end{matrix}
\right ]
\left[ \begin{matrix}{\bf a}_t \\ {\bf a}_w \\ {\bf v} \\ \bar {\bf a}_w\end{matrix} \right]_{k-1}
\end{equation}
More compactly:
\begin{equation}
{\bf a}'_k = {\bf T} {\bf a}'_{k-1}.
\end{equation}
Stability is assured if
\begin{equation}
\left | \lambda({\bf T}) \right | < 1
\end{equation}
that is the eigenvalues of ${\bf T}$ are all inside the unit circle.

Stability can be enforced if we do two things:

1) use a {\bf leaky integrator} for the actuators, i.e. replace the ${\bf I}$'s in the first term of \eqref{eq:controller} by $\gamma {\bf I}$, where $0< \gamma < 1$.

2) multiply the reconstructor matrix by a {\bf feedback gain}:
\begin{equation}
{\bf H}^\dagger \rightarrow g {\bf H}^\dagger
\end{equation}
where $g$ is made sufficiently small.

As $g \rightarrow 0$ the eigenvalues of ${\bf T}$ converge to three degenerate eigenvalues, $\gamma$, $\alpha$, and $\beta$ which are all less than $1$ in magnitude. Therefore there is a range of gains $g>0$ where the system is stable. The response time of the system to input disturbance is
\begin{equation}
\tau_r = - T/\ln \left| \lambda_{\rm max} \right|
\end{equation}
where $T$ is the sample period.

For further analysis it is instructive to note that only the mode sets defined through $A_t$ and $A_w$ are dynamically affected by feedback. The orthogonal parts of the Hilbert space are in the null space of the reconstructor, so they are neither excited by the disurbance nor fed back but are simply left to decay at a rate set by $\gamma$ without any affect on long-term stability. If we carry just the selected mode coefficients in our state-vector analysis, the stability  equation is:

\begin{equation}
\left[ \begin{matrix}{\bf c}_t \\ {\bf c}_w \\ {\bf v} \\ \bar {\bf c}_w \end{matrix}\right]_k =
\left[ \begin{matrix} \gamma{\bf I} - g {\bf H}_t^\dagger {\bf H}_t  & {\bf 0} &  {\bf M}_t {\bf C}_{wt}^T & - g {\bf H}_t^\dagger {\bf H}_w \\
                      -g {\bf M}_w {\bf C}_{wt} {\bf H}_t^\dagger {\bf H}_t    & \gamma{\bf I} & {\bf 0}  & -g {\bf M}_w {\bf C}_{wt} {\bf H}_t^\dagger {\bf H}_w \\
               -(1-\alpha) g {\bf M}_w {\bf C}_{wt} {\bf H}_t^\dagger {\bf H}_t & {\bf 0} & \alpha {\bf I} & -(1-\alpha) g {\bf M}_w {\bf C}_{wt} {\bf H}^\dagger {\bf H}_w \\
                          {\bf 0} & (1-\beta) {\bf I} & {\bf 0} & \beta {\bf I} \end{matrix}\right]
\left[ \begin{matrix}{\bf c}_t \\ {\bf c}_w \\ {\bf v} \\ \bar {\bf c}_w \end{matrix}\right]_{k-1}
\end{equation}

We now make an approximation and some simplifying assumptions to illucidate key points about closed loop behavior of the woofer-tweeter system.
First, assume that the reconstructor obeys
\begin{equation}
{\bf H}_t^\dagger {\bf H}_t \approx {\bf I}
\end{equation}
which is to say that all the tweeter modes are visible to the Hartmann sensor and that they are uniquely identifiable by it. If we strictly exclude invisible modes from the tweeter mode space and apply very little actuator penalty or a-priory statistic weighting, i.e. ${\bf Q}\approx{\bf 0}$ in \eqref{eq:H}, then this will be close to the case.

Next, let's arrange it so the modes of the woofer match exacly a subset of modes of the tweeter, and furthermore that  the modes in this set are orthonormal. This can be done in practice via a Gram-Schmidt process assuming, as stated earler, that each woofer mode can be closely fit by the tweeter. Then
\begin{equation}
{\bf C}_{wt} = \left[ \begin{matrix} {\bf I}_{n_w} & {\bf 0}\end{matrix} \right] \quad\quad {\bf H}_t^\dagger {\bf H}_w \approx \left[ \begin{matrix} {\bf I}_{n_w} \\ {\bf 0} \end{matrix} \right]
\end{equation}
and
\begin{equation}
{\bf M}_w = {\bf I}_{n_w} \quad\quad {\bf M}_t = {\bf I}_{n_t}
\end{equation}
where $n_w$ is the number of controlled modes on the woofer and $n_t$ is the number of controlled modes of the tweeter.
Then the stability equation is
\begin{equation}
\left[ \begin{matrix}{\bf c}_t \\ {\bf c}_w \\ {\bf v} \\ \bar {\bf c}_w \end{matrix}\right]_k =
\left[ \begin{matrix} (\gamma - g ) {\bf I} & {\bf 0} & \left[\begin{matrix} {\bf I}_{n_w} \\ {\bf 0} \end{matrix}\right] & - g \left[\begin{matrix} {\bf I}_{n_w} \\ {\bf 0}\end{matrix}\right] \\
                      -g \left[\begin{matrix} {\bf I}_{n_w} & {\bf 0}\end{matrix}\right]   & \gamma {\bf I} & {\bf 0}  & -g {\bf I}\\
               -(1-\alpha) g \left[\begin{matrix} {\bf I}_{n_w} & {\bf 0}\end{matrix}\right]  & {\bf 0} & \alpha {\bf I} & -(1-\alpha) g {\bf I}  \\
                          {\bf 0} & (1-\beta) {\bf I} & {\bf 0} & \beta {\bf I} \end{matrix}\right]
\left[ \begin{matrix}{\bf c}_t \\ {\bf c}_w \\ {\bf v} \\ \bar {\bf c}_w \end{matrix}\right]_{k-1}
\end{equation}
The dynamics separate into two independent subspaces, one associated with the modes shared by woofer and tweeter, and one associated with tweeter modes not being sent to the woofer. The equation for a shared mode is
\begin{equation} \label{eq:overlapDynamics}
\left[ \begin{matrix}c_{t \in w} \\ c_w \\ v\\ \bar c_w\end{matrix}\right]_k =
\left[ \begin{matrix} (\gamma- g) &  0 & -1 & - g \\
                            -g & \gamma & 0 & -g \\
                   -(1-\alpha) g  & 0 & \alpha & -(1-\alpha) g \\
                           0 & (1-\beta) & 0 & \beta \end{matrix} \right]
\left[ \begin{matrix}c_{t \in w} \\ c_w \\ v\\ \bar c_w\end{matrix}\right]_{k-1}
\end{equation}
and the equation for an isolated tweeter mode is
\begin{equation}
\left[ \begin{matrix}c_{t\notin w}\end{matrix}\right]_k = (\gamma - g) \left[ \begin{matrix}c_{t\notin w} \end{matrix}\right]_{k-1}
\end{equation}.

Each mode obeys an independent equation with 4-element state vectors for woofer-tweeter shared modes and a single-element state for modes isolated to the tweeter.

It is common and desirable to set the integrator leak, $\gamma$ near unity, and the gain $g$ also near unity. In that case the tweeter-isolated equations are clearly stable ($\left|\gamma-g\right| < 1$). In the coupled case, the 4x4 matrix is marginally stable; one eigenvalue is at $\lambda = +1$ with associated eigenvector $\left[ \begin{matrix} -1 & 1 & 0 & 1 \end{matrix} \right]^T$. This eigenmode corresponds to equal and opposite values on the woofer and tweeter integrators, cancelling each other. This ``woofer-tweeter fighting" mode remains at zero however, as the internal structure of the controller prevents external disturbances or noise from exciting it (see Figure~\ref{fig:blockDiagram}). Mathematically, the input vector, $\left[ \begin{matrix} g & g & (1-\alpha) g & 0 \end{matrix}\right]^T$, through which all disturbances must enter, is orthogonal to this eigenvector.

The remaining three eigenvalues are dominated by $\alpha$, $\beta$, and $\gamma-g$, that is, the time constant associated with the woofer-tweeter crossover, the time constant of the woofer response, and the closed-loop time response of the tweeter. The response time constant of ShaneAO's woofer (this is a magnetic voice-coil actuator membrane mirror manufactured by ALPAO SAS of Montbonnot, France) is 0.005 seconds (1/200 Hz) or, for a sample time of 1 ms, $\beta = 0.82$. If the low-pass filter coefficient $\alpha$ is chosen to match $\beta$, and $0<\gamma<1$ and $0<g<1$, then clearly all three remaing poles are well inside the unit circle and the system is stable. Simulations have demonstrated that the system is robust to a wide range of changes to the design parameters $\alpha$, $\gamma$, and $g$ within the range $[0,1)$.

To understand performance, and evaluate whether we reach our goal of putting most of the low-order mode excursion on the woofer, we consider  the closed-loop response to a step function disturbance at an overlapping mode. This mode enters the shared mode dynamics \eqref{eq:overlapDynamics}  via the input vector $\left[ \begin{matrix} g & g & (1-\alpha) g & 0 \end{matrix}\right]^T$. We conclude that, if $\gamma \approx 1$, woofer modes $c_w$ tend to the input steady-state value while the overlapping tweeter modes $c_{t \in w}$ tend to zero (the converged state-vector is near $\left[ \begin{matrix} 0 & 1 & 0 & 1 \end{matrix}\right]^T$). This is in fact the desired behavior for overlapping woofer-tweeter modes, that they eventually bleed off of the tweeter end up completely on the woofer.

\begin{figure}
   \begin{center}
   \begin{tabular}{c}
   \includegraphics[width=\columnwidth]{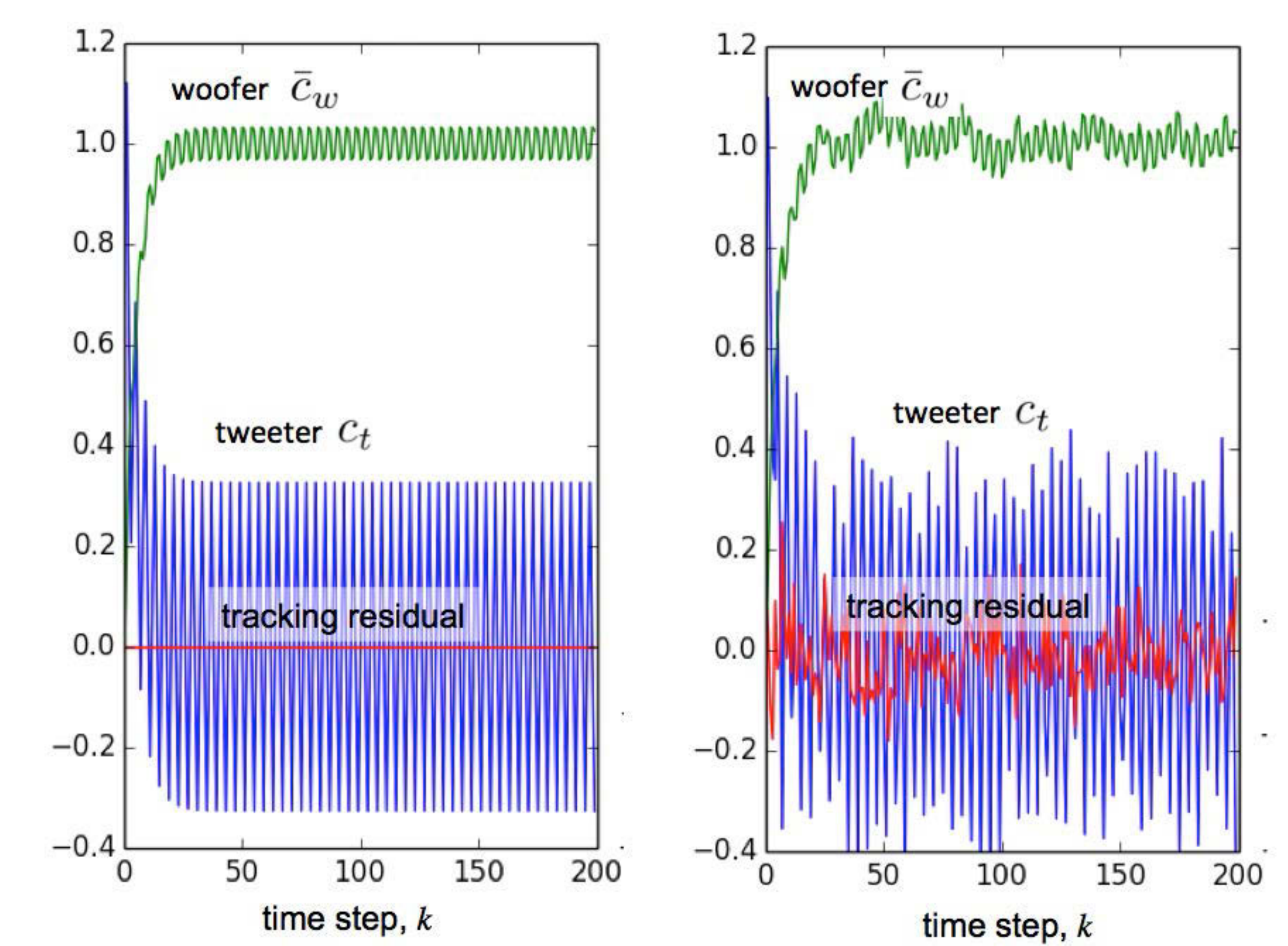}
\vspace{-2 mm}
   \end{tabular}
   \end{center}
   \caption[example] 
%>>>> use \label inside caption to get Fig. number with \ref{}
   { \label{fig:simulation} 
A simulation of the $c_{t\in w}$ and $\bar c_w$ states in response to disturbance of a unit step plus sinusoid of magnitude 0.3 at 250 Hz. Left: with zero measurement noise, right, with 0.07 rms measurement noise. The simulation parameters are $\alpha = 0.82$, $\beta = 0.82$, $\gamma = 1$, $g = 1$. As desired, the woofer controls the majority of the low-frequency part of the disturbance, the step function, while the tweeter controls most of the high frequency part of  disturbance, the sinusoid.} 
   \end{figure}

Figure \ref{fig:simulation} shows the results of simulating \eqref{eq:overlapDynamics} in response to a step function plus a sinusoid, and the same with added measurement noise. It is clear that the woofer eventually adjusts to match the step, while the higher frequency component is controlled mostly by the tweeter. The overall wavefront error is near zero from the start, showing that the tweeter's initial action makes up for the woofer's slow response.

%-------------
%-------------

%-------------
\begin{Video} \vspace{1 mm}
   \begin{center}
   \begin{tabular}{c}
   \includegraphics[height=2.2in]{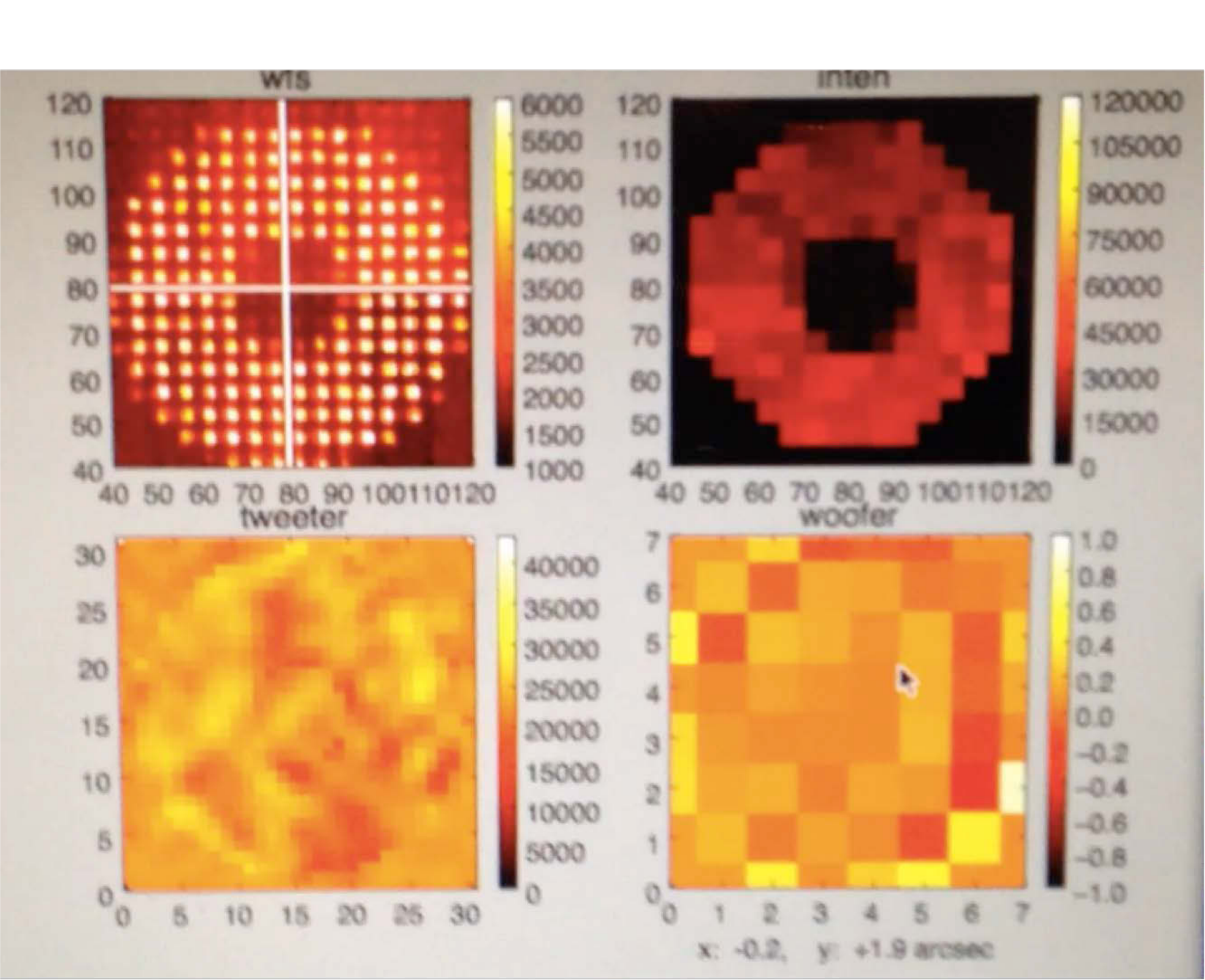}
   \end{tabular}
   \end{center}
   \caption[example] 
%>>>> use \label inside caption to get Fig. number with \ref{}
   { \label{vid:ShaneAOmovie} 
AO operator's display showing the wavefront sensor and commands to the woofer and tweeter deformable mirrors in the ShaneAO system. This video is taken live off the control screen during closed-loop AO operations with a natural guide star. http://dx.doi.org/doi.number.goes.here}
\end{Video}
%-------------

%%%%%%%%%%%%%%%%%%%%%%%%%%%%%%%%%%%%%%%%%%%%%%%%%%%%%%%%%%%%%
\section{CONCLUSION} 

This paper has presented a practical means for implenenting woofer-tweeter shared wavefront control in an adaptive optics system. Assumptions about the deformable mirrors and their working mode spaces are kept to a minimum in order to make it most generally applicable. One can use modal or zonal reconstructors, Fourier space, Zernike space, etc. within its context. Also, the reconstructor can incorposrate standard techniques for suppressing unobservable modes such as waffle or piston. The one key assumption made is that the chosen subset of woofer space should also be addressable by the tweeter. This allows two things: 1) the woofer space can be made orthogonal to the remaining part of the tweeter space; this prevents the woofer and tweeter (possibly unstably) working against each other, and 2) the tweeter can control the woofer space at temporal frequencies beyond the bandwidth of the woofer. The woofer-tweeter architecture employs mode splitting in the wavefront reconstruction, with a low-pass filter cross-over between woofer and tweeter. The closed-loop version of the algorithm is demonstrated to be stable and performs the desired goal of putting low-order components preferably on the woofer.

We are happy to report that the ShaneAO system, which uses this control architecture, is now commissioned on sky. The wavefront control is clearly demonstrating the desired closed-loop behavior with high-stroke control action going to the woofer and the tweeter keeping up with the high order aberrrations while rarely saturating.

%%%%%%%%%%%%%%%%%%%%%%%%

%\bibliography{Gavel_WooferTweeter_SPIE_2014}  %>>>> bibliography data in report.bib
%\bibliographystyle{spiebib}   %>>>> makes bibtex use spiebib.bst

%%%%%%%%%%%%%%%%%%%%%%%%%%%%%%%%%%%%%%%%%%%%%%%%%%%%%%%%%%%%%
\acknowledgments     %>>>> equivalent to \section*{ACKNOWLEDGMENTS}       
 
This research was funded in part by the National Science Foundation, Major Research Instrumentation grant \#0923585. The author gratfully acknowledges the support of the NSF and also the University of California Observatories for its cost-share and in-kind contributions that made the ShaneAO project possible and a success.  

\end{document}